\documentclass[12pt,preprint]{aastex}  
\shorttitle{The Prompt Gamma-Ray Emission of GRB~070125}
\shortauthors{Bellm et al.}

\newcommand{\Ep}{\ensuremath{E_{peak}}}

\begin{document}
 
\title{Observations of the Prompt Gamma-Ray Emission of GRB~070125}

\author{Eric C. Bellm\altaffilmark{1,2}, Kevin Hurley\altaffilmark{1},
Valentin Pal'shin\altaffilmark{3}, Kazutaka Yamaoka\altaffilmark{4}, 
Mark S. Bandstra\altaffilmark{1}, Steven E. Boggs\altaffilmark{1,5},
Soojing Hong\altaffilmark{6},
Natsuki Kodaka\altaffilmark{7},
A. S. Kozyrev\altaffilmark{8},
M. L. Litvak\altaffilmark{8},
I. G. Mitrofanov\altaffilmark{8},
Yujin E. Nakagawa\altaffilmark{9},
Masanori Ohno\altaffilmark{10},
Kaori Onda\altaffilmark{7}, 
A. B. Sanin\altaffilmark{8},
Satoshi Sugita\altaffilmark{4},
Makoto Tashiro\altaffilmark{7}, 
V. I. Tretyakov\altaffilmark{8},
Yuji Urata\altaffilmark{7}, 
 \and
Claudia Wigger\altaffilmark{11}
}
\altaffiltext{1}{UC Berkeley Space Sciences Laboratory, 7 Gauss Way,
Berkeley, CA 94720-7450, USA}
\altaffiltext{2}{ebellm@ssl.berkeley.edu}
\altaffiltext{3}{ Ioffe Physico-Technical Institute, 26 Polytekhnicheskaya,
St. Petersburg 194021, Russian Federation}
\altaffiltext{4}{Department of Physics and Mathematics, 
Aoyama Gakuin University, 5-10-1 Fuchinobe, Sagamihara, 
Kanagawa 229-8558, Japan}
\altaffiltext{5}{Department of Physics, University of California, Berkeley}
\altaffiltext{6}{Laboratory of Physics, College of Science and Technology, 
Nihon University, 7-24-1 Narashinodai, Funabashi, Chiba 274-8501, Japan}
\altaffiltext{7}{Department of Physics, Saitama University,
255 Shimo-Ohkubo, Sakura-ku, Saitama, Saitama 338-8570, Japan}
\altaffiltext{8}{Institute for Space Research, Profsojuznaja 84/32,
Moscow 117997, Russia}
\altaffiltext{9}{Institute of Physical and Chemical Research (RIKEN), 2-1
Hirosawa, Wako, Saitama 351-0198, Japan}
\altaffiltext{10}{Institute of Space and Astronautical Science, Japan
Aerospace Exploration Agency (ISAS/JAXA), 3-1-1 Yoshinodai, Sagamihara, 
Kanagawa 229-8510, Japan}
\altaffiltext{11}{Paul Scherrer Institute, Villigen PSI, Switzerland}

\begin{abstract}
The long, bright gamma-ray burst GRB~070125 was localized by the 
Interplanetary Network. We present light curves of the prompt gamma-ray
emission as observed by Konus-WIND, RHESSI, Suzaku-WAM, and
\textit{Swift}-BAT.
We detail the results of joint spectral fits with Konus and RHESSI data.
The burst shows moderate hard-to-soft evolution in its multi-peaked
emission over a period of about one minute.  The total burst fluence as
observed by Konus is $1.79 \times 10^{-4}$~erg/cm$^2$ (20~keV--10~MeV).  
Using the spectroscopic
redshift $z=1.548$, we find that the burst is consistent with the
``Amati'' $E_{peak,i}-E_{iso}$ correlation.  Assuming a jet opening angle
derived from broadband modeling of the burst afterglow, GRB~070125 is a 
significant outlier 
to the ``Ghirlanda'' $E_{peak,i}-E_\gamma$ correlation.  Its
collimation-corrected energy release $E_\gamma = 2.5 \times 10^{52}$~ergs 
is the largest yet observed.

\end{abstract}

\keywords{gamma-rays: bursts} 

%%%%%%%%%%%%%%%%%%%%%%%%%%%%%%%%%%%%%%%%%%%%%%%%%%%%%%%%%%%%%%%%%%%%%%%%%%%%
\section{Introduction} 

The prompt gamma-ray emission of gamma-ray bursts (GRBs) is the most
extensively studied aspect of these energetic explosions.  Indeed, for
twenty-five years after the discovery of GRBs \citep{kleb73}, the prompt
emission was the only GRB observable available.  With the
first afterglow observations at longer wavelengths \citep{cost97,vanp97},
detailed analysis of burst models became possible.  
Presently, the \textit{Swift} satellite is detecting $\sim$ 100 bursts
per year, most with rapid localization and followup.

The exact mechanism which produces the prompt gamma-ray emission, with its
characteristic smoothly broken power-law spectrum, has not been definitively
established.  Recent efforts to correlate burst
observables with the intrinsic burst energetics have increased the
importance of detailed spectral fitting for localized bursts
\citep[for a review, see][]{zhan07c}.
Some correlations involve the peak spectral energy \Ep,
which is often above the $\sim$150 keV cutoff of the \textit{Swift}
Burst Alert Telescope (BAT) passband.

Several current observatories are capable of detailed spectral analysis of
GRBs over the full range of \Ep.  Konus-W \citep{apte95} 
is a double scintillator instrument on the
WIND spacecraft.  The Ramaty High Energy Solar Spectroscopic Imager
(RHESSI) is a solar observatory which uses
nine germanium detectors to image the Sun at X-ray to gamma-ray energies
\citep{lin02}.  RHESSI's detectors are unshielded and receive emission
from astrophysical sources like GRBs.  The Wide-Band All-Sky Monitor (WAM)
\citep{yama05} aboard Suzaku is the large BGO anticoincidence shield for the 
Suzaku Hard X-Ray Detector.  AGILE \citep{trav06}
and GLAST \citep{ritz07} will give
additional coverage at the energy range of \Ep\ and extend
spectral coverage for GRBs up to tens of GeV.

In this paper, we present Konus, RHESSI, and Suzaku observations of the
bright GRB~070125.  In Section \ref{sec-obs}, we discuss the observations and
the localization of the burst by the IPN.  Section \ref{sec-lc} contains
the burst light curves, and in Section \ref{sec-fits} we conduct joint
spectral fits to the Konus and RHESSI data.

%%%%%%%%%%%%%%%%%%%%%%%%%%%%%%%%%%%%%%%%%%%%%%%%%%%%%%%%%%%%%%%%%%%%%%%%%%%%

\section{Observations} \label{sec-obs} 

GRB~070125 was observed by six spacecraft in the Interplanetary Network
(IPN):  RHESSI, Suzaku WAM, and \textit{Swift}-BAT, all in
low Earth orbit; the anticoincidence system of the spectrometer aboard
the International Gamma-Ray Astrophysics Laboratory (INTEGRAL), at 0.44
light-seconds from Earth; Konus-Wind, at 5.4 light-seconds from Earth;
and the High Energy Neutron Detector and Gamma Sensor Head aboard Mars
Odyssey 2001, at 1130 light-seconds from Earth.  The two other distant
missions in the network, Ulysses and MESSENGER (Mercury Surface, Space
Environment, Geochemistry, and Ranging), were off.  Since \textit{Swift} was
slewing at the time of the burst, it did not immediately localize it.
However, the source appeared in a routine image made after the slew was
completed, and its 2.5' radius error circle was consistent with the
initial IPN localization \citep{igcn070125}.  Even with more than six
minutes of elapsed time since the burst onset, the BAT image detections were
highly significant at 8.2 sigma \citep{racu07}.

\begin{figure}
\plotone{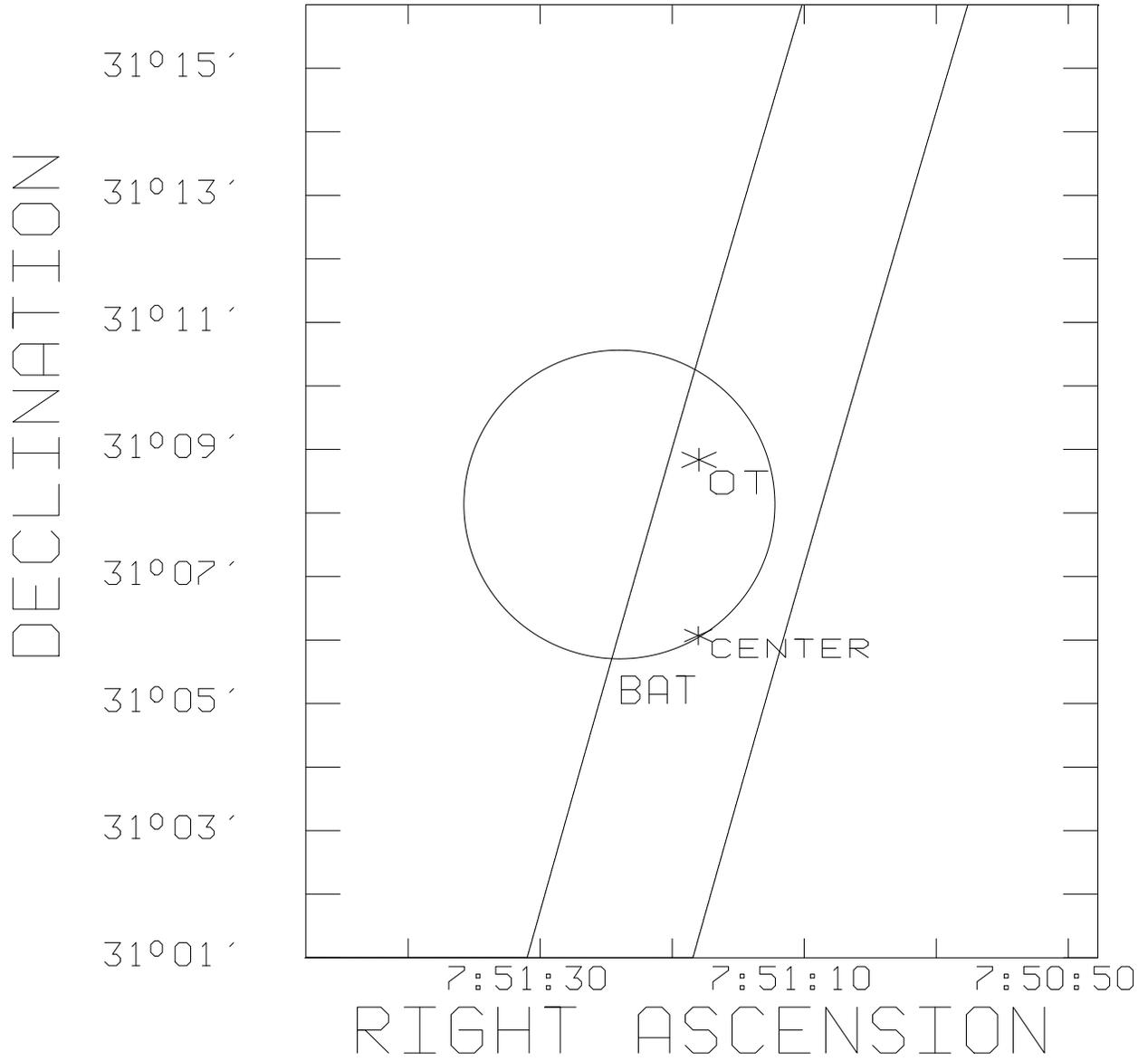}
\caption{The central region of the IPN error ellipse, showing
the 2.5' radius BAT $3\sigma$ error circle, the optical
transient source first reported by \citet{cenk07}, and the
center of the ellipse.  The optical source lies 0.048 degrees
from the center of the IPN ellipse, on the 87\% confidence contour.}
 \label{fig-zoom}
\end{figure}

With only one distant spacecraft, the IPN localized the event
to a long, narrow error ellipse 
whose area ($3\sigma$) is $\sim$ 1200 square arcminutes, centered at 
RA(2000) = 07h 51m 17.85s,  Dec(2000) = +31$^\circ$ 06' 12.78".  
The chi-squared for this
position is 1.57 for 3 degrees of freedom.  Figure \ref{fig-zoom} shows
the central region of the error ellipse, with the BAT 90\% confidence
error circle and the optical counterpart.  

Initial spectral fits to the prompt emission were reported for RHESSI by
\citet{rgcn070125} and for Konus by \citet{kgcn070125}.  The initial RHESSI
best fit model was a cutoff power law (equivalent to the Band function below $E_{break}$, see \S \ref{sec-fits}) with $\alpha$
= 1.33 $^{+0.11}_{-0.09}$, 
\Ep = 980. $\pm$ 300.~keV, and a 30 keV--10~MeV fluence of 
$1.5 \times 10^{-4}$~erg/cm$^2$.  The Konus data were best fit by a 
Band function with $\alpha$ =
-1.10 $^{+0.10}_{-0.09}$, $\beta$ = -2.08 $^{+0.10}_{-0.15}$, 
\Ep\ = 367 $^{+65}_{-51}$~keV.  The measured Konus
20 keV--10~MeV fluence was $(1.74 ^{+0.18}_{-0.15}) \times 10^{-4}$ 
erg/cm$^2$.  All errors are 90\% C.L.

Pelangeon and Atteia derived a pseudo-redshift for this burst by using
the RHESSI parameters \citep{pela07a} and the Konus values
\citep{pela07b}.  These were fairly consistent at 1.6 $\pm$ 0.8 and 
1.3 $\pm$ 0.3 respectively.

\citet{cenk07} reported an optical counterpart at RA(2000)=07h~51m~17.75s, 
Dec(2000)~=~+31$^\circ$~09'~04.2".  This counterpart was confirmed
by \citet{updi07a} in the R band. 

\citet{racu07b} reported detection by the \textit{Swift} XRT.
The XRT position was RA(J2000)  = 7h~51m~18.08s, 
Dec(J2000) = +31$^\circ$~09'~02.2", 4.7 arcseconds from the optical
transient reported by \citet{cenk07}.

Initial afterglow detections in other bands included 
\textit{Swift} UVOT in the UV \citep{mars07b}, radio \citep{vand07}, 
and IR \citep{bloo07a}.  Milagro \citep{ding07} observations 
of the source took place, but no VHE gamma-ray source was detected. 

\citet{fox07} reported a redshift of $z \geq 1.547$ for GRB~070125 from the 
identification of the Mg II doublet.  \citet{cenk08} tightened this
estimate to $z = 1.5477 \pm 0.0001$.  Independent observations by
\citet{proc07}, reported by \citet{updi08},
reveal absorption features which are consistent with $z = 1.548$
if identified as C IV and Si IV, and the absence of Lyman absorption
features requires $z$ to be near this value.

Observations of the decaying afterglow yielded multiple possibilities for a
jet break.  The \textit{Swift}-XRT data showed a
possible jet break at 1.35 $\pm$ 0.35 days, but were also
consistent with no jet break \citep{burr07,racu07}.  Independent optical
observations \citep{mira07,garn07} showed a break in the decay at $t \geq
4$ days.  The non-detection by Chandra \citep{cenk07b} was also consistent
with a break occuring after 4 days.  \citet{updi08} used a larger optical 
dataset to fit a jet break time of $t = 3.73 \pm 0.52$ days, but cautioned
that flaring made the best fit break time dependent on the choice of time
intervals.  \citet{chan08} found a best fit break time of $t = 3.8$ days 
in a joint optical--X-ray fit.  They suggested that the break
might be chromatic, as the X-ray data alone did not require a break, and
proposed that inverse Compton emission could create a delay between the
optical and X-ray breaks.

Extensive observations of the afterglow of GRB~070125 allowed detailed
studies of the unusual burst environment.
\citet{cenk08} suggested that the low absorbing column densities 
inferred from the afterglow spectra indicate that this long burst took
place in a low-density galactic halo.  \citet{chan08} performed detailed
broadband fitting of the afterglow, and concluded that the immediate
environment of the progenitor was likely high density ($n \sim 50$ cm$^{-3}$
for a constant density profile).  They also found evidence that the
gamma-ray production efficiency for this burst was unusually high
($\eta_\gamma \sim 0.65$).

%%%%%%%%%%%%%%%%%%%%%%%%%%%%%%%%%%%%%%%%%%%%%%%%%%%%%%%%%%%%%%%%%%%%%%%%%%%%
\section{Light Curve} \label{sec-lc}

\begin{figure} 
\plotone{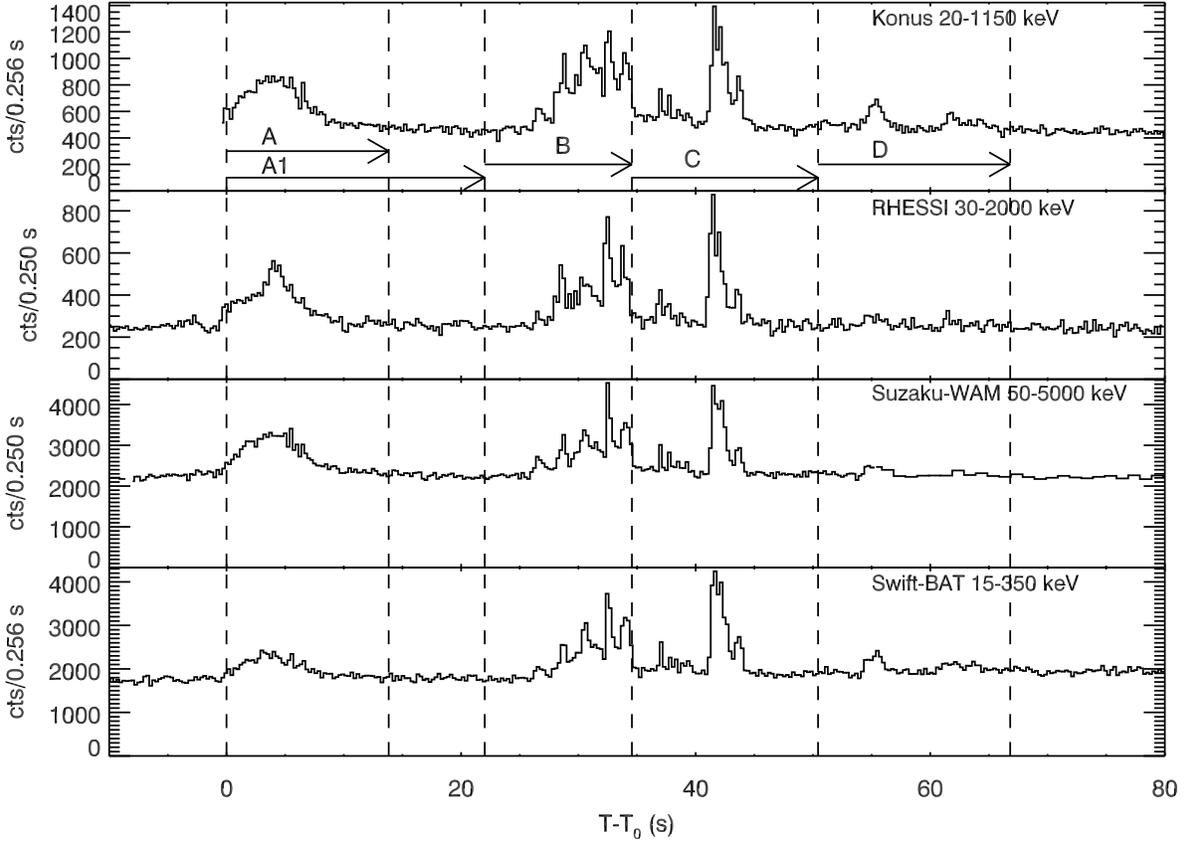}
\caption{GRB~070125 light curve for Konus, RHESSI (rear
segments 1, 7, and 8 only), Suzaku-WAM, and \textit{Swift}-BAT.  The
	light curves are adjusted for time of flight, with T$_0$ given
		in \S \ref{sec-lc}.  The dashed vertical lines delimit the
		intervals used in the time-resolved spectral fits (\S \ref{sec-fits}).
The \textit{Swift} light curve plotted contains all counts observed by
\textit{Swift}; in particular, it is not mask-tagged and therefore contains
slew artifacts.  
}
\label{fig-lc}
\end{figure}

Figure \ref{fig-lc} shows the Konus, RHESSI, Suzaku-WAM, and
\textit{Swift}-BAT light curves
corrected for light travel time between the spacecraft.  The Konus trigger
time was T$_{0,KW}$ = 07:20:50.853.  Photon travel time
from RHESSI to Konus was 5.197 seconds, from Suzaku to Konus was
5.202 seconds, and from \textit{Swift} to Konus was 5.215 seconds.  

The light curves show a qualitatively similar 
multi-peaked structure with roughly four major periods of emission.  
The RHESSI data in interval A have a slight but significant feature around
$T_0$ + 4 seconds whose origin is unclear.  The bump appears in data from
all three detectors used in this study.  Examination of hardness ratios
suggests that the bump is softer than the rest of the emission in the
interval, but insignificantly so ($\sim 1 \sigma$).  The difference is even
more negligable when we consider only data above 65~keV.  Accordingly, the
bump (if extraneous) should not meaningfully influence the spectral fits 
reported in Section \ref{sec-fits}.

T90 for the Konus light curve was 62.2 $\pm$ 0.8 seconds (20--1150~keV), 
for RHESSI 63.0 $\pm$ 1.7 seconds (30 keV--2~MeV), and
for Suzaku 55 $\pm$ 2 seconds (50 keV--5~MeV).  
In the individual Konus bands, the T90s were 62.8 $\pm$ 1.8 seconds 
(G1: 20~keV--75~keV), 61.5 $\pm$ 0.9 seconds (G2: 75~keV--300~keV), 
and 60.0 $\pm$ 5.6 seconds (G3: 300~keV--1150~keV).  
Uncertainties on all T90s are 1-sigma and
were obtained by perturbing the light curves with Poisson noise and finding
the new T90 values for 1000 trials.
\citet{racu07} report a T90 of 60 seconds for the
\textit{Swift}-BAT light curve.  Because \textit{Swift} did not trigger
on the burst, no BAT event data were stored. The available rate data contain
slew artifacts; accordingly, we do not perform further analysis on the BAT
data.

Both Konus and RHESSI observed the 64 millisecond peak flux at T-T$_0 =$ 41.472
seconds.  Using the spectral fits from Section \ref{sec-fits}, the peak
flux (20~keV--10~MeV)
observed by Konus was $(1.85 ^{+0.35}_{-0.36}) \times 10^{-5}$ 
erg/cm$^2$/s.  RHESSI observed a peak flux of $(2.92 ^{+0.68}_{-0.63}) 
\times 10^{-5}$ erg/cm$^2$/s.  
While the RHESSI fluences computed in Section
\ref{sec-fits} are lower than those measured by Konus, RHESSI recorded a
greater proportion of counts in the 64 ms peak interval, implying a 
larger peak flux.  These values are moderately sensitive to background
subtraction; the errors quoted are purely statistical.

\begin{figure} 
\plotone{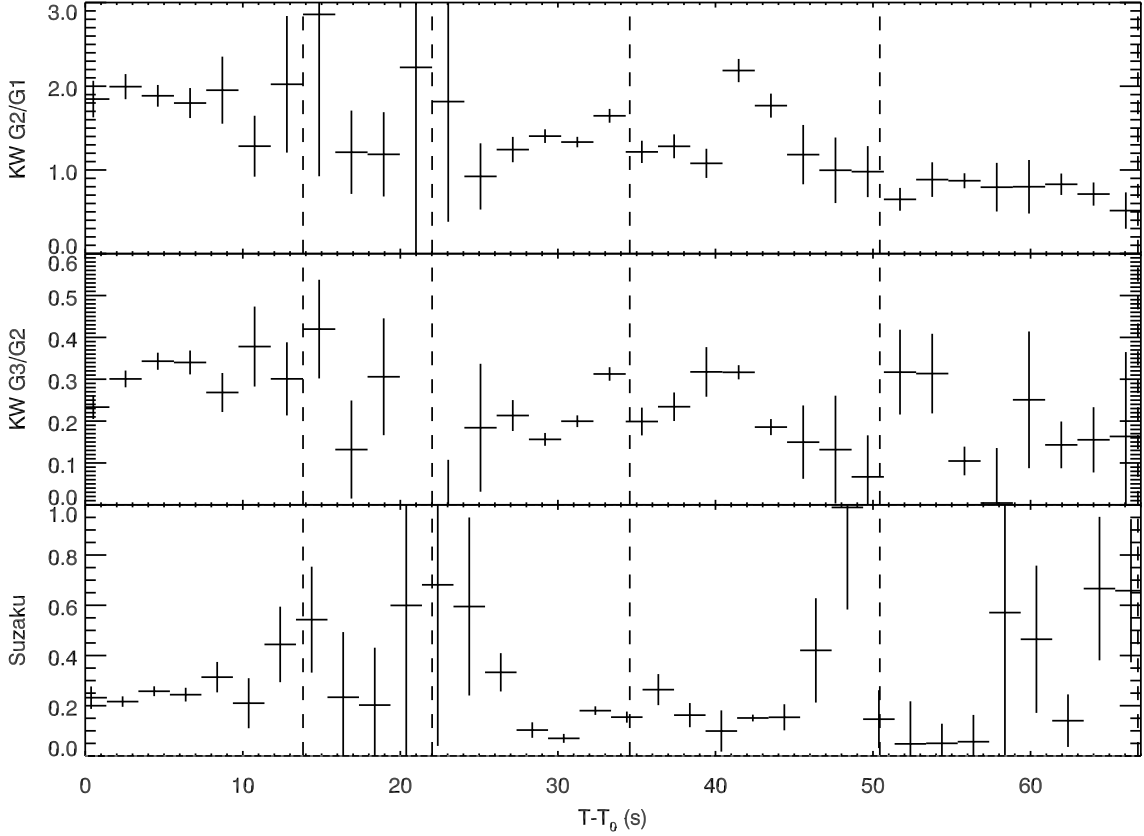}
\caption{Hardness ratios for GRB~070125.  The Konus energy bands are G1
(20--75~keV), G2 (75--300~keV), and G3 (300--1150~keV).
The Suzaku hardness ratio plotted here is (520--5000~keV)/(50--240~keV).  
%Hardness
%ratios of less than zero indicate imperfect background subtraction.  
Dashed
lines indicate spectral fitting intervals, as in Figure \ref{fig-lc}.
Points near $\sim$20 seconds 
which are off-scale for the Konus G3/G2 ratio are consistent with
zero---there is negligible emission in the G3 band at this time.
}
\label{fig-hardness}
\end{figure}

Figure \ref{fig-hardness} shows the fast time evolution of hardness ratios for 
Konus and Suzaku.  The burst shows a general softening trend in time,
excepting the period of peak flux in interval C, which has comparable
hardness to the initial emission in interval A.

%%%%%%%%%%%%%%%%%%%%%%%%%%%%%%%%%%%%%%%%%%%%%%%%%%%%%%%%%%%%%%%%%%%%%%%%%%%%
\section{Spectral Analysis} \label{sec-fits}

We performed spectral analysis for the time intervals given in Table
\ref{tab-intervals} using the Konus and RHESSI data.   While spectral
data are available from Suzaku, the GRB photons passed through the X-Ray
Spectrometer (XRS) dewar before reaching the WAM.  This direction is not
well-calibrated for the WAM, in part due to uncertain levels of solid Ne in
the dewar.  With the detector response poorly understood, it is impossible
to determine effectively the spectral parameters.  Accordingly, we
omit the Suzaku data in the spectral fits.

Konus 64-channel spectra are available beginning 0.512
seconds before the trigger and are integrated over variable timescales.  
The detector
response, which is a function only of the burst angle relative to the 
instrument axis, is
generated from Monte Carlo simulations described by \citet{tere98}.

Because of radiation damage to the RHESSI detectors, only three of the
nine detectors (rear segments 1, 7, and 8) were usable for this analysis.
While the damaged detectors continue to record significant counts, the
effect of the radiation damage on the spectral response has proven difficult
to model.

To generate the RHESSI spectral response, we simulated monoenergetic
photon beams impinging on a detailed mass model in the Monte Carlo suite
MGEANT \citep{stur00}.  
The response of each detector changes as RHESSI rotates, so we
used a beam geometry with photons generated along 60$^\circ$ arcs in rotation
angle.  The resulting sector responses were weighted by the burst light
curve and added together. Fit results were not appreciably different when
using a simple azimuthally averaged response.
The beam made an angle of 165$^\circ$ with the RHESSI rotation axis to
match the off-axis angle of the GRB (165.2 degrees).  The simulated photons 
had initial energies given by 192 logarithmically-spaced bins from 10~keV to 
30~MeV. 

We conducted the spectral fitting in parallel using the spectral fitting
packages XSPEC v11\footnote{http://heasarc.gsfc.nasa.gov/docs/xanadu/xspec/}
and ISIS v1.4.3 \citep{citeisis}.  The fit
parameters obtained from both programs were identical.  Robust fitting
required a lower fit bound of 65 keV for RHESSI, slightly higher than the
typical 30 keV lower limit.  Because the GRB was arriving from the extreme
rear of RHESSI, the photons passed through the back plate of the RHESSI 
cryostat and were hence subject to greater attenuation at low energies.
The fit ranges were accordingly 20~keV--10~MeV for Konus and 65~keV--10~MeV
for RHESSI.   We rebinned the data to a minimum S/N of 2 before performing
the spectral fits.  This rebinning did not greatly affect the best fit
parameters.
Fluence errors were obtained in ISIS by stepping through a grid of
fluence values, refitting the free parameters at each grid point, and
monitoring the change in chi-squared.  Since it does not assume that the
statistic space is quadratic, this method provides more accurate
values for the uncertainties than those generated in XSPEC with the
\texttt{flux} command.

\begin{figure}
\plotone{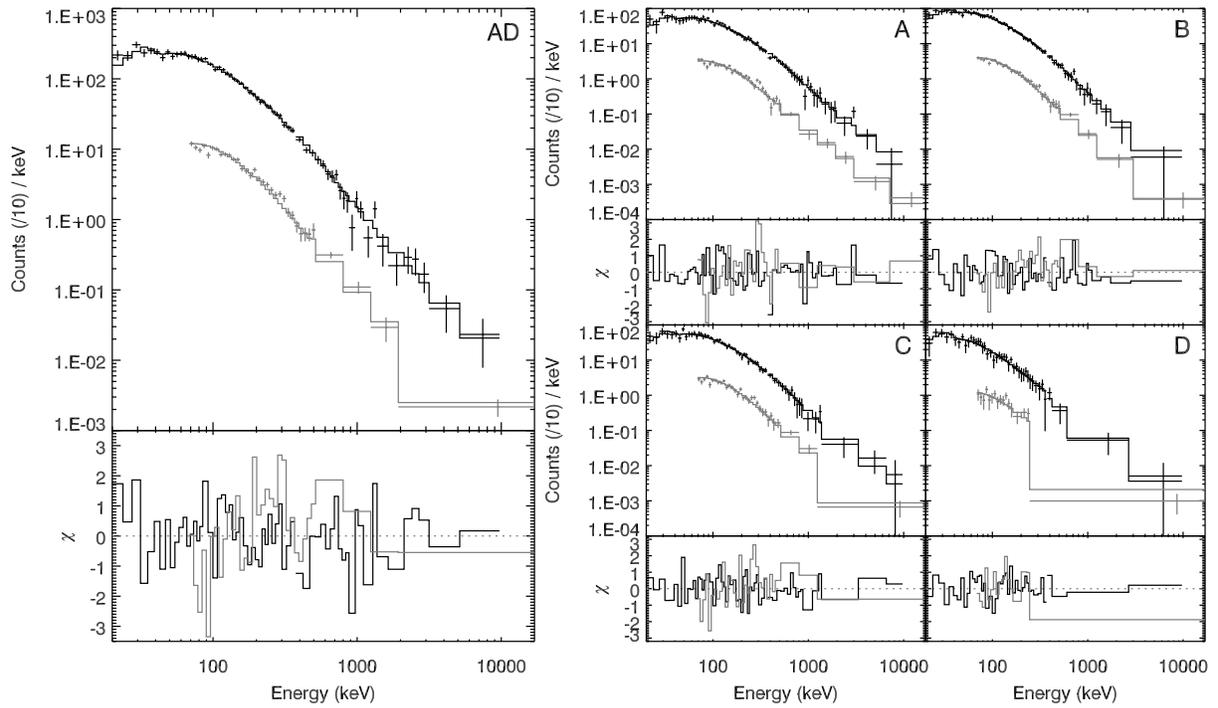}
\caption{Count spectra and residuals for the joint fits.  
The Konus data and models are colored black, while the RHESSI data 
and models are gray.
RHESSI data, model, and errors are divided by 10 
in the count spectra plots for clarity.
The overplot models differ only in normalization.}
\label{fig-spec}
\end{figure}

The data were well-fit in intervals A-C by a Band function \citep{band93}:
\[
N_E = \left\{
\begin{array}{lr}
  A (E/E_{piv})^{\alpha} \exp(-E (2+\alpha)/\Ep) & E<E_{break} \\
  B	(E/E_{piv})^{\beta}               & E>E_{break}
\end{array}
\right.
\]
with $E_{break} \equiv \Ep \frac{(\alpha - \beta)}{(2+\alpha)}$ and 
$B \equiv  A (\frac{(\alpha-\beta) \Ep}{(2+\alpha) E_{piv}})^{\alpha-\beta}
\exp(\beta-\alpha)$.  For $\beta < -2$ and
%$\alpha > -2$, $\Ep \equiv E_0 (2 + \alpha)$ corresponds to the peak of the
$\alpha > -2$, \Ep\ corresponds to the peak of the
$\nu F_\nu$ spectrum.  The normalization $A$ has units 
photons/(cm$^2$~s~keV), and $E_{piv}$ is here taken to be 100~keV. 
For joint fits, the Band function parameters $\alpha$,
$\beta$, and \Ep\ were tied for both instruments,
but the normalizations were allowed to vary independently.
For interval D, the best fit model after grouping was a simple
power law.
%($E_{piv}$ was 1 keV for the cutoff power law fit.
We report the best-fit spectral parameters in Table \ref{tab-fitpars}.
Figure \ref{fig-spec} shows the spectra in all intervals for the
joint fit. 

For single-instrument fits, the Konus data provide superior fit quality
and better constraint on the fit parameters, due in part to having about
six times more usable counts.  The fit fluence, $\alpha$,
and $\beta$ are generally consistent between RHESSI and Konus.  However,
the RHESSI data prefer higher \Ep, matching the best fit Konus values 
only at the lowest end of rather large error bars.  
The Konus fit parameters for the total burst match well the initial values 
reported via the GCN \citep{kgcn070125}.  The RHESSI fit \Ep\ typically  is lower here
than in the value reported in the GCN \citep{rgcn070125}, but this
difference is expected from fitting using the Band function rather than a
cutoff power law \citep{band93}.

The spectral parameters for the joint fits are consistent with
the Konus-only values.  There are slight improvements
in the uncertainties of some of the fit parameters at a cost of an increase
in the chi squared.  The RHESSI residuals in the joint fit (Figure
\ref{fig-spec}) show a characteristic deviation pattern, indicating that
the instruments disagree on the spectral shape.  The Konus data dominate
the fit because of their better statistical quality.  The residuals
for the RHESSI-only fits do not show any systematic deviation.

For intervals A, and B, the ratio of the RHESSI normalization to the
Konus normalization is 0.88.  For interval C, the ratio is 0.95.
Characteristic uncertainties for the ratio are 0.04-0.05.  
In interval D, the ratio for the power-law fit is
0.84$_{-0.13}^{+0.14}$.
Absolute normalizations in photons/(cm$^2$~s~keV) 
using $E_{piv} = 100$ keV for the total interval were 
(2.50 $^{+0.18}_{-0.15}) \times 10^{-2}$ (Konus) and 
(2.25 $^{+0.16}_{-0.14}) \times 10^{-2}$ (RHESSI).

The time-resolved fits show a moderate hard-to-soft evolution.  \Ep\ is
largest in the initial broad pulse (539 keV) and then softens to 355 keV in
interval B.  The sharp pulse in interval C has a harder spectrum (418 keV).  
While the statistically preferred model for the S/N grouped data in interval D 
is a simple power
law, fitting a cutoff power-law with the Konus data to 2~MeV gives an
estimate of \Ep\ at 220~keV.
The high-energy spectral index $\beta$ softens monotonically through intervals 
A-C.

%%%%%%%%%%%%%%%%%%%%%%%%%%%%%%%%%%%%%%%%%%%%%%%%%%%%%%%%%%%%%%%%%%%%%%%%%%%%
\section{Energetics}

Knowledge of the burst redshift $z=1.548$ makes it possible to draw
conclusions about the overall burst energetics.
We assume a standard flat cold dark matter cosmology
($\Lambda$CDM), with parameters ($\Omega_\Lambda$, $\Omega_M$, $H_0$) 
= (0.761, 0.239, 73~km~s$^{-1}$~Mpc$^{-1}$), 
consistent with results from WMAP year 3 \citep{sper07}
and large scale structure traced by luminous red galaxies \citep{tegm06}.
This particular set of values corresponds to the ``Vanilla model'' of
\citet{tegm06}.

\begin{figure}
\plotone{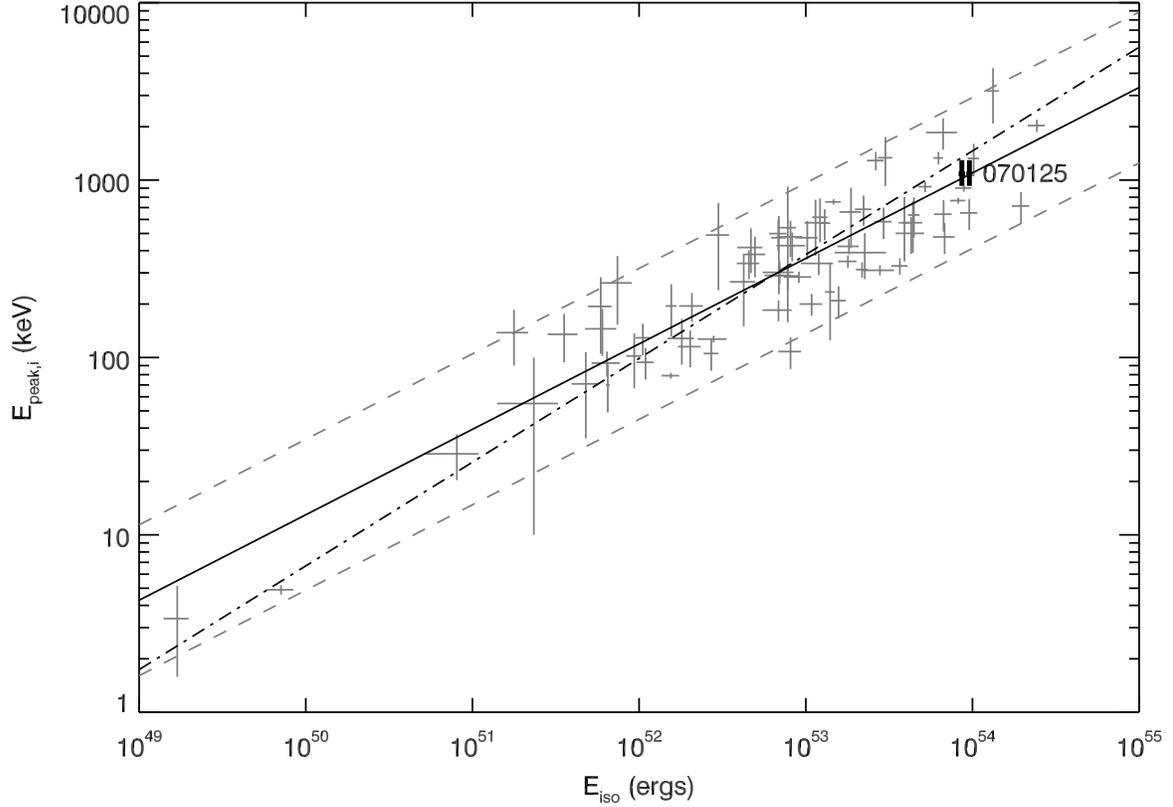}
\caption{$E_{peak,i} - E_{iso}$ correlation including GRB~070125.  Values of
$E_{peak,i}$ (the intrinsic peak energy in the burst rest frame)
and $E_{iso}$ are for the joint Konus-RHESSI fit.  Since the
normalization was allowed to vary between the two instruments, we plot
separate points for Konus and RHESSI to
indicate the corresponding values of $E_{iso}$.  The Konus data point has the 
larger value of $E_{iso}$.  Data for other bursts are
from Table 1 of \citet{ghir08}, 
plotted using the cosmology of this paper ($\Omega_m = 0.239$, $\Omega_\Lambda
= 0.761$, $h=0.730$).  
The best-fit line for the unweighted data points, omitting GRB~070125, 
is overplot with a solid line; the $2\sigma$ scatter about that fit is 
indicated with dashed lines.  The dash-dotted line is the best fit when
the data points are weighted by their errors on both axes, again 
omitting GRB~070125---see text for details.
}
\label{fig-amati}
\end{figure}

Extrapolating to a GRB rest-frame
energy band of 1~keV--10~MeV, the isotropic emitted energy for the total
burst is $(9.59\pm0.39) \times 10^{53}$~ergs (Konus) and 
$(8.67\pm0.38) \times 10^{53}$~ergs (RHESSI) for the joint fit.  
Because we allow independent normalizations for the Konus and RHESSI data,
we obtain two values of $E_{iso}$ from the joint fit, one for each
instrument.
90\% C.L. errors are obtained by exploration of the parameter space as for
the fluence; we neglect uncertainty in $z$.  These values, together with
the spectral fit of the time-integrated spectrum, are
consistent with the ``Amati relation'' correlating $E_{iso}$ with the intrinsic
peak energy of the spectrum in the GRB rest frame $E_{peak,i}$
\citep{amat02,amat06,ghir08}.  We plot GRB~070125 
in the $E_{peak,i} - E_{iso}$ plane in Figure \ref{fig-amati}.

Because the best fit Band function has a hard tail ($\beta \sim -2$), the
fluence integral is sensitive to the choice of upper energy bound.
If we use the observed energy band 20~keV--10~MeV,
corresponding to a GRB frame band of 50~keV--25.5~MeV, 
the fluence is 14\% larger than that in the usual bolometric band.
For consistency with previous works, we
will use the 1~keV--10~MeV band for bolometric estimates.

Converting the 64~ms peak fluxes reported in Section \ref{sec-lc} to
bolometric peak luminosities using the best fit Band parameters,
we find peak luminosities of (2.59
$^{+0.36}_{-0.37}) \times 10^{53}$~ergs/s for Konus and 
(4.25 $^{+0.87}_{-0.79}) \times 10^{53}$~ergs/s for RHESSI.

\begin{figure}
\plotone{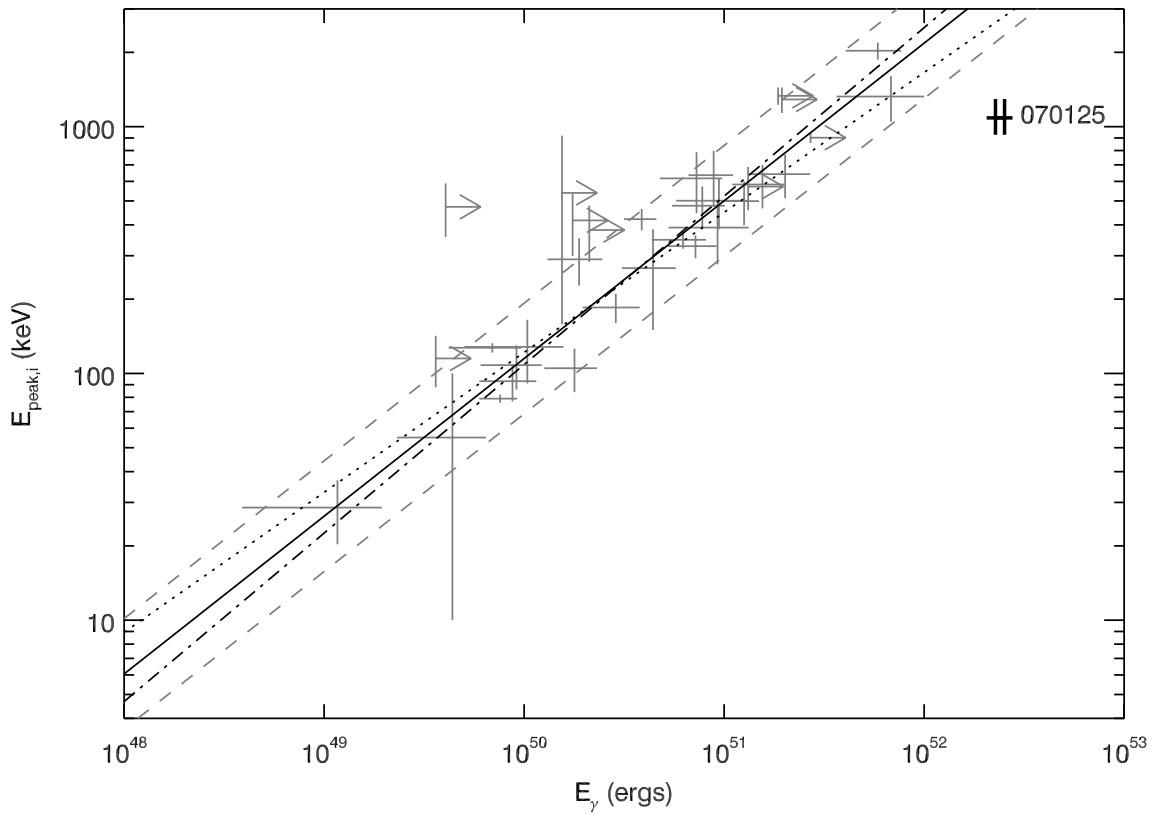}
\caption{$E_{peak,i} - E_{\gamma}$ correlation including GRB~070125. 
Symbols and overplot fit lines are as in Figure \ref{fig-amati}; the Konus data
point has the larger value of $E_{\gamma}$.  We also plot the best unweighted
fit line including GRB~070125 with a short dotted line.
Data for other bursts are from Table 1 of \citet{ghir07}, 
assuming an ISM density profile and plotted using the cosmology of this paper. 
Bursts with only lower limits on $E_{\gamma}$ were omitted from the fit.
}
\label{fig-ghirlanda}
\end{figure}

\citet{chan08} performed a broadband fit to afterglow data for GRB~070125.
They determined a jet opening angle of $13.2 \pm 0.6$ degrees in their most
plausible scenario (a radiative fireball expanding into a constant density 
(ISM) medium and emitting via synchrotron and inverse Compton channels).
This jet angle was consistent with that inferred from the jet break time
$\sim 3.7$ days and an emission radius derived from radio scintillation.  
For the collimation-corrected energy
$E_\gamma = (1 - \cos \theta) E_{iso}$, we find $E_\gamma = 
(2.52 \pm 0.24, 2.27 \pm 0.22) \times 10^{52}$ ergs for (Konus, RHESSI).
These values are the largest yet recorded for a burst with measured \Ep\ 
(c.f. \citealp{frai06}; \citealp{koce08} also reported lower limits on
$E_{\gamma}$ greater than $10^{52}$ ergs for several \textit{Swift} bursts 
using the time of the last XRT observation).
%\citep[cf.][]{frai06}.
We plot GRB~070125 
in the $E_{peak,i} - E_{\gamma}$ plane in Figure \ref{fig-ghirlanda} to examine
its consistency with the ``Ghirlanda'' $E_{peak,i} - E_\gamma$ correlation 
\citep{ghir04,ghir07}. 

In Figures \ref{fig-amati} and \ref{fig-ghirlanda}, we also overplot the best
fit correlation lines.  A number of fitting approaches have been considered 
in the literature in an effort to account for the apparent
extra-statistical spread of the points about the 
correlation \citep[for a review, see][]{ghir08}.  We have followed
\citet{ghir08} in presenting two least squares fits, one in which the data
points are unweighted and a second in which the errors on both axes are
considered.  After the fit, we estimate the dispersion of the points
perpendicular to the best fit correlation line
using the square root of the bias-corrected sample variance.

GRB~070125 is quite consistent with the Amati relation:
including it in the fit makes negligible changes in the best-fit correlation
slope or the logarithmic dispersion ($0.20$ dex).  However, it is a
$5.0\sigma$ outlier to the Ghirlanda correlation fitted without it, using the sample
dispersion to estimate $\sigma$.  
Including GRB~070125 in an unweighted fit of the bursts in the Ghirlanda
sample, the overall dispersion increases to 0.13~dex (from 0.09~dex), and 
GRB~070125 remains a $2.8\sigma$ outlier.

The unusual environment of GRB~070125 is responsible for its high value of
$E_{\gamma}$.  In particular, the jet opening angle of $13.2 \pm 0.6$
degrees derived by \citet{chan08} is larger than all of those
presented by \citet{ghir07}.  
Retaining the $3.7 \pm 0.5$ day jet break time well-established
in the optical \citep{updi08,chan08}, we may derive the jet opening angle
assuming adiabatic emission and more conventional parameters \citep{sari99}.
Assuming an ISM profile with circumburst density $n = 3$ cm$^{-3}$ and a
gamma-ray production efficiency of $\eta_\gamma = 0.2$, the corresponding jet
opening angle is $\theta = 5.6 \pm 0.3$ degrees for 
Konus.  The resulting collimation-corrected energy would be $E_{\gamma} = (4.6
\pm 0.5) \times 10^{51}$~ergs, only $0.8\sigma$ from the best-fit correlation
omitting GRB~070125.

\section{Discussion}

While GRB~070125 had a large measured prompt gamma-ray fluence, its spectral
properties are unremarkable.  The values of the best-fit spectral parameters
are similar to those observed for other bright bursts
\citep[e.g.,][]{kane06}, and
the spectral evolution observed is similarly common.  The environment
of GRB~070125 is unique, however \citep{cenk08, chan08, updi08}, requiring
a broad jet opening angle in broadband afterglow models \citep{chan08}.
After collimation correction, GRB~070125 has the most energetic prompt
emission yet observed and is a significant outlier to the correlation between
peak energy and $E_{\gamma}$.

%Granting the existance of these correlations, however,
GRB~070125 appears to
weaken the claim that the Ghirlanda correlation has low dispersion.  
GRB~070125 is not a ``recognizable'' outlier to the Ghirlanda relation
in the sense of \citet{ghir07}, as
it is highly consistent with the Amati relation.
Its jet parameters have been derived from a rich and well-sampled afterglow
dataset.  While the circumburst environment of this GRB is unusually dense,
this only highlights the assumption of a fairly narrow range of efficiency and
density parameters for the majority of GRBs where broadband modeling of the
afterglow has not been possible.  The true dispersion of the correlation may in
fact be larger.  

%The physical significance of the GRB luminosity correlations has been 
%questioned (e.g., \citealp{butl07,butl08}; but see \citealp{ghir08}).  
%Selection effects due to
%detector thresholds influence the sample, and transforming parameters to the
%GRB rest frame can create illusory correlations.
The physical significance of GRB spectrum--energy correlations has been
questioned \citep[e.g.][]{butl07,butl08}.  In particular, 
detector trigger thresholds affect burst detection, and more complex
selection effects govern the measurement of peak energies,
redshifts, and afterglow breaks.  These effects can influence the sample of
GRBs with known redshift, $E_{peak,i}$, and $E_{\gamma}$.  \citet{ghir08}
examined the effect of trigger and spectral analysis thresholds in the
\Ep--fluence plane, finding that the \textit{Swift}-detected burst sample
was truncated by the spectral analysis threshold.  Neither threshold 
truncated the pre-\textit{Swift} burst sample.  
%and the authors speculate that other selection effects determine the
%extent of the pre-\textit{Swift} sample.

We were unable to confirm the source of the systematic shift
in \Ep\ and fluence between the two instruments for this burst.
Minor radiation damage was becoming noticeable in
RHESSI detector 8 near the time of this work, mostly below the 65 keV cut
utilized here.  
It is also possible that the Monte Carlo simulation
of the RHESSI response is less accurate for such extreme off-axis angles,
where a greater number of interactions with the cryostat may be expected.  

%\citet{krim06} found for a cutoff power-law
%fit to combined \textit{Swift} and Konus data for GRB 050717 a best fit
%value of \Ep$=2401 +781/-568$.  A joint \textit{Swift}-RHESSI fit to the
%same burst found \Ep$=1950 \pm 350$ keV \citep{bellhead06}.  
Our previous work had found excellent agreement in all fit parameters
for independent RHESSI and Konus spectral fits for GRB 051103 and 
GRB 050717.  For the short GRB
051103, Konus found $\Ep = 1920 \pm 400$~keV and a 20~keV--10~MeV fluence of
$4.4 \pm 0.5 \times 10^{-5}$~ergs/cm$^2$ \citep{kgcn051103, fred07}.  A RHESSI
fit yielded $\Ep = 1930 \pm 340$~keV and 20~keV--10~MeV fluence of $4.5 \times
10^{-5}$~erg/cm$^2$ \citep{bellhead06}.
\citet{krim06} found for a cutoff power-law
fit to Konus data for GRB 050717 a best fit
value of \Ep $=2101 ^{+1934}_{-830}$~keV.  A RHESSI fit to the same burst found
\Ep $=1550 ^{+510}_{-370}$~keV \citep{wigg06}.  
Those bursts had RHESSI off-axis angles of 97 and 110 degrees, respectively.

Joint spectral fits to \textit{Swift}-BAT and RHESSI data for 25 bursts
co-observed by the two instruments between December 2004 and December 2006
indicated that no offset in response normalization was needed for the two
instruments \citep{bellsf08}.
However, for two of three bursts occurring during or after December 2006, 
the RHESSI data showed a significant deficit relative to \textit{Swift}-BAT.  
The RHESSI polar angles for all three late bursts were between 90 and 110 
degrees.
These fits were conducted using only detectors 1 and 7, which do not appear
to have radiation damage in background spectra during this interval.  
Nonetheless, these results suggest that the observed offset in the RHESSI and 
Konus fit parameters found here is more likely a consequence of increased
radiation damage in the RHESSI detectors than a geometric effect or a
generic offset in the RHESSI simulations.

Future analysis of archival bursts may help identify the source of any
systematic effects present here.  It is clear, however, that joint fits
between instruments capable of constraining the full range of \Ep\ are
valuable in providing the most accurate and precise determination of the fit
parameters.

\acknowledgements 
This work was supported by Swift AO-2 GI grant NNG06GH58G, ``Completing
Swift GRB Energy Spectra with Konus and RHESSI'' and by the A0-3 grant 
NNX07AE86G.  KH is grateful for IPN support under JPL Contract 1282043, and 
NASA grants NNG06GI896, NNX06AI36G, NNG06GE69G, and NAG5-13080.  
The Konus-Wind experiment is supported by a Russian Space Agency 
contract and RFBR grant 06-02-16070.
We thank Bob Lin, David Smith, and Dieter Hartmann for helpful comments.

{\it Facilities:} \facility{RHESSI}, \facility{WIND (Konus)}

\bibliography{grb}
\bibliographystyle{apj}

\clearpage

%----- Table of column densities at tau=1 -----%
\begin{deluxetable}{lrrr}
\tablecolumns{4}
\tablewidth{0pt}

\tablecaption{\label{tab-intervals}
Intervals used for spectral fitting in \S \ref{sec-fits}.  The reference
time for Konus is T$_{0,KW}$ = 07:20:50.853.  For RHESSI, T$_{0,R}$ =
T$_{0,KW}$ - 5.197 s.
}

\tablehead{\colhead{Interval} & \colhead{T$_i$ - T$_0$ (s)} & 
								\colhead{T$_f$ - T$_0$ (s)} & 
								\colhead{T$_f$ - T$_i$ (s)} 
}

\startdata
A  &  0    & 13.824 &13.824	\\
A1 &  0    & 22.016 &22.016 \\
B  & 22.016 &  34.560 &12.544 \\
C  & 34.560 &  50.432 &15.872 \\
D  & 50.432 & 66.816 &16.384 \\
\tableline
AD & 0      & 66.816 & 66.816 \\
\enddata

\end{deluxetable}
 
\begin{deluxetable}{llllcr}
\tablecolumns{6}
\tablewidth{0pt}
\rotate

\tablecaption{\label{tab-fitpars}
Best fit parameters for the Band function for Konus (K) and RHESSI (R).
Errors are quoted at the 90\%
confidence level.  For joint fits (KR), the Konus fluence is listed first.  For
interval D, the fit and quoted fluence are for a power-law model. 
}

\tablehead{\colhead{Instruments} & 
	\colhead{$\alpha$} & \colhead{$\beta$} &
	\colhead{\Ep} & \colhead{20 keV-10 MeV Fluence} & \colhead{$\chi^2$/dof} \\

	\colhead{} &\colhead{} &\colhead{} &\colhead{(keV)} &
		\colhead{(10$^{-5}$ erg/cm$^2$)} &\colhead{} 
}

\startdata

\cutinhead{Total Burst (Intervals A1-D)}
K & -1.09$^{+0.09}_{-0.08}$ & -2.09$^{+0.10}_{-0.15}$ &   373$^{  +66}_{  -51}$ & 17.2$^{+1.5}_{-1.5}$ &  63/ 60 =  1.05 \\
R & -0.90$^{+0.46}_{-0.28}$ & -2.24$^{+0.20}_{-0.44}$ &   533$^{ +261}_{ -171}$ & 16.2$^{+1.7}_{-1.7}$ &  38/ 30 =  1.30 \\
KR & -1.13$^{+0.09}_{-0.08}$ & -2.08$^{+0.09}_{-0.14}$ &   430$^{  +80}_{  -61}$ & 17.9$^{+1.3}_{-1.3}$ & 130/ 93 =  1.40 \\
 & & & & 16.1$^{+1.1}_{-1.1}$ & \\

\cutinhead{Interval A}
K & -0.89$^{+0.18}_{-0.15}$ & -1.99$^{+0.12}_{-0.23}$ &   447$^{ +154}_{  -99}$ &  6.20$^{+0.71}_{-0.75}$ &  65/ 61 =  1.08 \\
R & -0.52$^{+0.68}_{-0.42}$ & -2.12$^{+0.18}_{-0.30}$ &   512$^{ +249}_{ -157}$ &  5.49$^{+0.69}_{-0.71}$ &  45/ 32 =  1.43 \\
KR & -0.96$^{+0.14}_{-0.11}$ & -2.04$^{+0.12}_{-0.17}$ &   539$^{ +129}_{ -113}$ &  6.22$^{+0.56}_{-0.55}$ & 123/ 96 =  1.28 \\
 & & & &  5.48$^{+0.47}_{-0.47}$ & \\

\cutinhead{Interval B}
K & -1.07$^{+0.14}_{-0.11}$ & -2.21$^{+0.14}_{-0.23}$ &   318$^{  +65}_{  -55}$ &  4.88$^{+0.54}_{-0.53}$ &  45/ 54 =  0.84 \\
R & -1.12$^{+0.60}_{-0.28}$ & -2.33$^{+0.30}_{-1.20}$ &   556$^{ +362}_{ -236}$ &  4.82$^{+0.68}_{-0.71}$ &  18/ 30 =  0.60 \\
KR & -1.11$^{+0.12}_{-0.11}$ & -2.17$^{+0.12}_{-0.20}$ &   355$^{  +78}_{  -56}$ &  5.18$^{+0.45}_{-0.47}$ &  76/ 87 =  0.88 \\
 & & & &  4.61$^{+0.40}_{-0.41}$ & \\

\cutinhead{Interval C}
K & -0.98$^{+0.18}_{-0.14}$ & -2.19$^{+0.19}_{-0.42}$ &   360$^{  +98}_{  -74}$ &  4.09$^{+0.64}_{-0.66}$ &  44/ 54 =  0.82 \\
R & -0.92$^{+0.80}_{-0.34}$ & -2.61$^{+0.54}_{ 0.54}$ &   623$^{ +450}_{ -287}$ &  4.02$^{+0.78}_{-0.68}$ &  30/ 28 =  1.08 \\
KR & -1.03$^{+0.16}_{-0.13}$ & -2.19$^{+0.17}_{-0.35}$ &   418$^{ +113}_{  -84}$ &  4.28$^{+0.54}_{-0.56}$ &  87/ 85 =  1.03 \\
 & & & &  4.05$^{+0.50}_{-0.51}$ & \\

\cutinhead{Interval D}
K & -1.90$^{+0.07}_{-0.08}$ &  -  &  -  &  1.73$^{+0.28}_{-0.24}$ &  32/ 42 =  0.77 \\
R & -1.97$^{+0.20}_{-0.25}$ &  -  &  -  &  1.34$^{+0.40}_{-0.29}$ &  15/ 15 =  1.03 \\
KR & -1.87$^{+0.08}_{-0.08}$ &  -  &  -  &  1.84$^{+0.36}_{-0.29}$ &  46/ 57 =  0.81 \\
 & & & &  1.48$^{+0.26}_{-0.24}$ & \\

\enddata

\end{deluxetable}
 %\newpage

%\caption{The IPN $3\sigma$ confidence error ellipse derived from a six 
%spacecraft triangulation of GRB~070125.  The asterisk indicates the center of
%the ellipse.}

\end{document}